\documentclass[conference]{IEEEtran}
\IEEEoverridecommandlockouts
\usepackage{cite}
\usepackage{amsmath,amssymb,amsfonts}
\usepackage{algorithmic}
\usepackage{graphicx}
\usepackage{caption}
\usepackage{subcaption}
\usepackage{textcomp}
\usepackage{xcolor}
\usepackage{multirow}

\def\BibTeX{{\rm B\kern-.05em{\sc i\kern-.025em b}\kern-.08em
    T\kern-.1667em\lower.7ex\hbox{E}\kern-.125emX}}
     \usepackage{pgfplots}
\pgfplotsset{compat=newest}
\usepackage{tikz}
\usetikzlibrary{plotmarks}
\usetikzlibrary{arrows.meta}
\usepgfplotslibrary{patchplots}
\usepackage{grffile}

\newtheorem{defn}{Definition}[section]
\newcommand{\argmax}{\mathrm{argmax}}
    
\begin{document}

\title{ Location-Aided Beamforming in Mobile  Millimeter-Wave Networks\\}

\author{\IEEEauthorblockN{Sara Khosravi\IEEEauthorrefmark{1},
Hossein S.Ghadikolaei\IEEEauthorrefmark{2}, Jens Zander\IEEEauthorrefmark{1}, and Marina  Petrova \IEEEauthorrefmark{1}\IEEEauthorrefmark{3}}
\IEEEauthorblockA{ \IEEEauthorrefmark{1}School of EECS, KTH Royal Institute of the Technology, Stockholm, Sweden,\\  \IEEEauthorrefmark{2} Ericsson Research, Sweden, \IEEEauthorrefmark{3} RWTH Aachen University, Germany \\
Email: \{sarakhos, jenz, petrovam\} @kth.se, hossein.shokri.ghadikolaei@ericsson.com
}}

\maketitle

\begin{abstract}
Due to the large bandwidth available, millimeter-Wave (mmWave) bands are considered a viable opportunity to significantly increase the data rate in cellular and wireless networks. Nevertheless, the need for beamforming and directional communication between the transmitter and the receiver increases the complexity of the channel estimation and link establishment phase.   
Location-aided beamforming approaches have the potential to enable fast link establishment in mmWave networks. However, these are often very sensitive to location errors. In this work, we propose a beamforming algorithm based on tracking spatial correlation of the available strong paths between the transmitter and the receiver. We show that our method is robust to uncertainty in location information, i.e., location error and can provide a reliable connection to a moving user along a trajectory. The numerical results show that our approach outperforms benchmarks on various levels of error in the location information accuracy. The gain is more prominent in high location error scenarios.  
\end{abstract}


\section{Introduction}

Millimetre-wave (mmWave) communication is one of the important components of fifth-generation cellular networks due to the large bandwidth available between 24 GHz and \mbox{86 GHz} frequency range  \cite{de2017making,ghatak2020beamwidth}. Nevertheless, due to the high penetration loss and fewer scattering paths, mmWave signals are vulnerable to blockage. To compensate for the severe loss and establish and maintain a reliable connection, both transmitter (Tx) and receiver (Rx) require directional communication using a large number of antennas and beamforming. The short wavelength of mmWave transmissions enables the integration of a large number of antennas and applying beamforming techniques on both Tx and Rx sides in many applications. 
Moreover, directional communication reduces the interference of other active base stations (BSs) and increases the data rates \cite{rappaport2013millimeter}. 

On the downside, the use of directional communication requires additional computations for beamforming tasks, especially in mobile scenarios where mobility causes frequent misalignment between Tx and Rx beams and consequently frequent re-execution of the beamforming procedure. The additional overhead for link maintenance reduces the average transmission rate of a user. Therefore, there is a tradeoff between beamforming overhead as a function of a number of re-execution of beamforming tasks and the instantaneous achieved rate of the user. 

Many of the existing mmWave beamforming approaches are based on exhaustive search procedures over a set of pre-defined beam directions in order to find the optimal alignments between a Tx and an Rx \cite{5174147,6171799}. However, these approaches increase the overhead due to the high dimension of the beam-searching space when using multi-antenna arrays with narrow beamwidth. Other approaches like sparsity-aware beamforming \cite{ghauch2016subspace} and compressive-sensing based beamforming approaches \cite{Hassanieh} suffer from the overhead in the mobile scenarios or may have hardware implementation constraints, respectively. Authors in \cite{sur2016beamspy,8057188} utilized the sparsity and the spatial correlation of mmWave channels in adjacent locations and proposed their beam-searching methods based on that. However, the applicability of their methods in real mobile scenarios may be limited because they are designed for only stationary users \cite{sur2016beamspy} or increase the complexity of the beam-searching \cite{8057188}.

Side information aided or context-aware approaches use the location information to reduce the effective beam search space during the beam alignment phase and also to facilitate the handover decision during the handover phase. The work in \cite{ghatak2020beamwidth} established a trade-off between localization accuracy and communication performance and proposed a localization bound-assisted initial beam-selection method. The authors in \cite{scaling} showed the importance of location information in scaling mmWave networks to dense environments. The authors in \cite{ali2017millimeter,va2015beam} found the location information as the useful side information to assist the link establishment in mmWave communication networks. The authors in \cite{garcia2016location} first employed a positioning algorithm in order to localize the users relative to the BS and then used the position and orientation information to select the proper beam to initiate the data transmission.  The authors in \cite{maschietti2017robust} considered noisy location information and presented a beam alignment optimization scheme. In particular, the authors proposed a two-step beam pre-selection at the Tx and the Rx. Under similar conditions as \cite{maschietti2017robust}, authors in \cite{ igbafe2019location} proposed a beam alignment method where the Tx and the Rx jointly select the beams which outperforms the two-step method. In this work, we consider the noisy location information and proposed a beamforming method that provides a reliable connection along the trajectory.

Today's localization algorithms are primarily based on geometric localization where the location of the BS is known and the user location is determined based
on geometric constraints such as the distance and physical angular orientations between the BS and the user \cite{kanhere2021position}. Localization techniques such as time-of-arrival (ToA),
time-difference-of-arrival, and angle-of-arrival (AoA)
are based on measuring the distance or angle of a user
with respect to multiple BSs with known positions for line-of-sight (LoS) propagation. However, in indoor or outdoor non-LoS (NLoS) environments, due to the different angels and the larger delays of arriving multi-paths,
these techniques correspond to some localization error variance. For instance, the mean error of 10 m reported in \cite{kanhere2018position} with AoA localization technique based on NLoS indoor office measurements. Other techniques like using positioning reference signals which is a pseudo-random sequence sent by LTE BSs \cite{3gpp}, can provide about \mbox{15 m} raw resolution \cite{kanhere2018position}. We refer readers to \cite [Table III]{del2017survey} for more details regarding the performance of different localization methods in cellular systems.

In this paper, we propose a framework for using the location information as the input of the beamforming algorithm and consider the different degrees of uncertainties on this information. We assume there exists a location algorithm with a certain error model and propose a beamforming algorithm that is robust to the localization errors.
 We leverage the sparsity and spatial correlation of mmWave channels in adjacent locations and proposed a beamforming method. Inspired by the work in \cite{sur2016beamspy}, we define a path skeleton set as a sparse set of strongest available paths between the Tx and the Rx.  Our proposed beamforming method is based on searching over path skeleton sets not all the directions. We explore and analyse the tradeoffs between i) location information accuracy and the achieved transmission rate, and ii) antenna beamwidth and robustness toward location information error. Unlike our prior works in \cite{Sara,Sara2}, the emphasis of this work lies in accounting for the uncertainties in the location information.

The rest of this paper is organized as follows. We introduce the system model in Section \ref{system model} and present our beamforming algorithm in Section \ref{BF}. The localization error model and beamforming with locution error are discussed in Section \ref{BF with e}. We present the numerical results in Section \ref{simulations}, and conclude our work in Section \ref{conclusions}.

\textit{Notations:} Throughout the paper, matrices, vectors and scalars are denoted by bold upper-case ($\mathbf{X}$), bold lower-case ($\mathbf{x}$) and non-bold ($x$ or $X$) letters, respectively. The $\ell_{2}$-norm and conjugate transpose  of a vector $\mathbf{x}$ (or a matrix $\mathbf{X}$) are represented by $\|\mathbf{x}\|$ and $\mathbf{x}^H$, respectively. We define set $[M]:=\{1,2,..,M\}$ for any integer $M$.

\section{System Model} \label{system model}
\subsection{Network Model}
We consider a downlink transmission scenario, where a Tx with $N_{\text{Tx}}$ antennas establishes communication with a single Rx/user with $N_{\text{Rx}}$ antennas\footnote{Our proposed methods are applicable to the uplink as well.}. In our scenario, the Rx is moving along a trajectory with length $M$.
We present the blockage model in Section \ref{simulations}.

We employ a narrow band channel model with a sparse number of dominant paths. In this model the channel matrix  $\mathbf{H}\in\mathbb{C}^{N_{\text{Rx}}\times N_{\text{Tx}}}$  between Tx and the Rx in location $i$ during a coherence interval \footnote{ The channel response during the coherence interval is approximately unchanged.} is defined as \cite{akdeniz2014millimeter}:

\begin{equation}
 \mathbf{H}_i=\sqrt{\frac{N_{\text{Tx}}N_{\text{Rx}}}{L}}\sum_{\ell=1}^{L} h_{i,\ell}\mathbf{a}(\phi^{\text{Rx}}_{i,\ell},\theta^{\text{Rx}}_{i,\ell}) \mathbf{a}^{H}(\phi^{\text{Tx}}_{i,\ell},\theta^{\text{Tx}}_{i,\ell}),
\label{H}
\end{equation}
where $L$ is the number of paths. Each path $\ell$ has the horizontal and vertical angle of arrivals (AoAs), $\phi^{\text{Rx}}_{i,\ell}$, $\theta^{\text{Rx}}_{i,\ell}$, and horizontal and vertical angle of departures (AoDs), \mbox{$\phi^{\text{Tx}}_{i,\ell}$, $\theta^{\text{Tx}}_{i,\ell}$,} respectively.
$h_{i,\ell} \sim \mathcal{N}_\mathbb{c} (0, \beta_{i,\ell})$ is the small scale fading, where $\beta_{i,\ell}$ is the channel gain. Here, $\mathbf{a}(\cdot)$ is the antenna array response. In this work, we consider a half wavelength uniform planar arrays (UPA) in yz-plane both in the Tx and the Rx sides. The Tx antenna array response is:
\begin{equation} \small
\mathbf{a}(\phi^{\text{Tx}}_{i,\ell},\theta^{\text{Tx}}_{i,\ell})=\frac{1}{\sqrt{N_{\text{Tx}}}}[1, ..., e^{j\pi [n_x \sin(\theta^\text{Tx}_{i,\ell})\sin(\phi^\text{Tx}_{i,\ell})+n_y \cos(\phi^\text{Tx}_{i,\ell})]},...]^T,
\end{equation}
where $1 \leq n_x \leq N_x-1$, $1 \leq n_y \leq N_y-1$ and $N_x$ and $N_y$ are the number of columns and rows of the UPA and $N_{\text{Tx}}=N_x\times N_y$.
\begin{figure}[t]
\centering
		\includegraphics[width=0.42\textwidth]{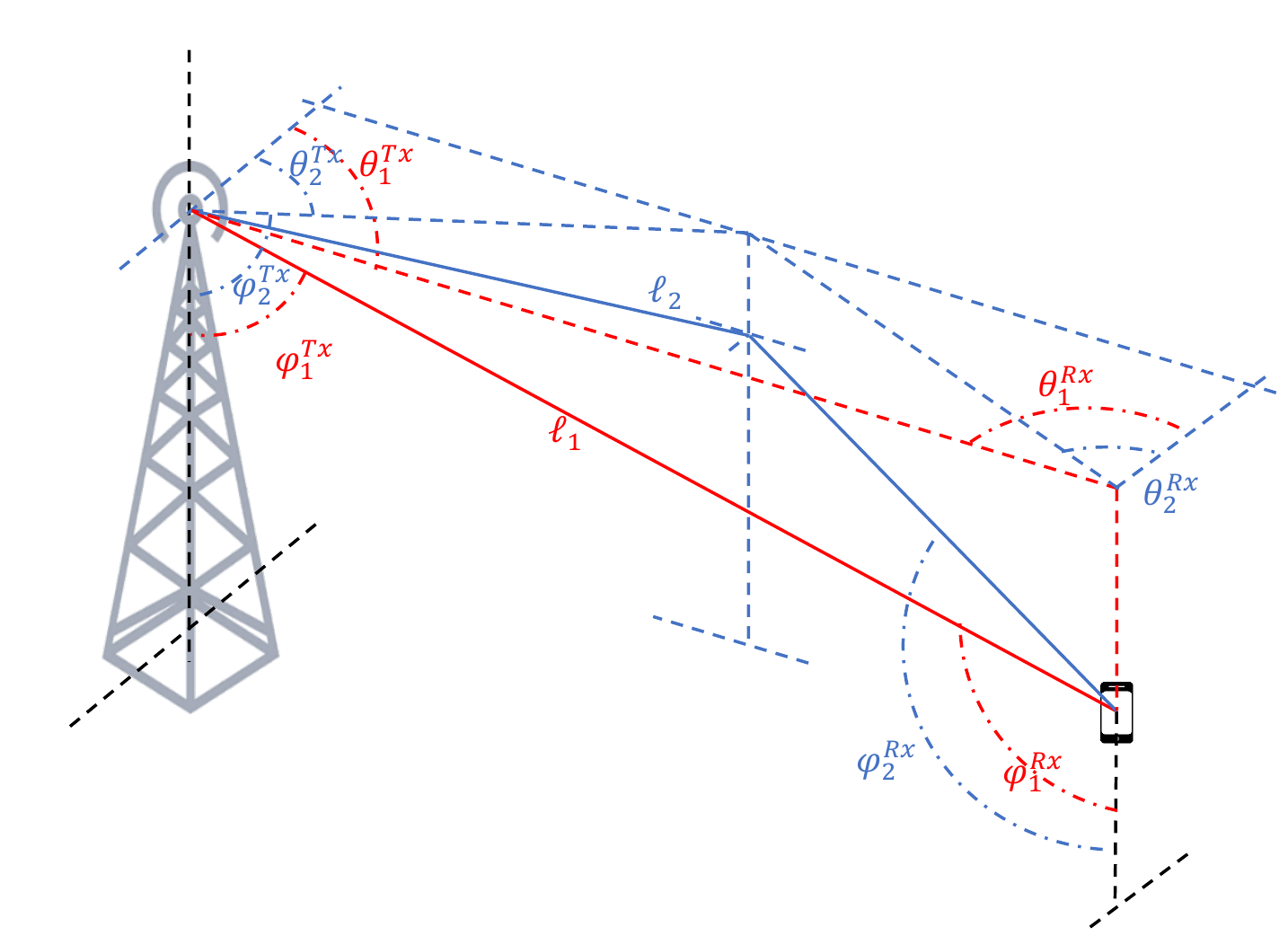}
		\caption{A path skeleton set, contains a LoS path
($\ell_1$) and a NLoS path ($\ell_2$)}
\label{ps}
	\end{figure}

The Rx antenna array response is defined as
\begin{equation}\small
\mathbf{a}(\phi^{\text{Rx}}_{i,\ell},\theta^{\text{Rx}}_{i,\ell})=\frac{1}{\sqrt{N_{\text{Rx}}}}[1, ..., e^{j\pi[m_x \sin(\theta^\text{Rx}_{i,\ell})\sin(\phi^\text{Rx}_{i,\ell})+m_y \cos(\phi^\text{Rx}_{i,\ell})]},...]^T,
\end{equation}
where $1 \leq m_x \leq M_x-1$, $1 \leq m_y \leq M_y-1$ and $M_x$ and $M_y$ are the number of columns and rows of the UPA and $N_{\text{Rx}}=M_x\times M_y$.

\subsection{Beamforming Model}
We consider analogue beamforming at both the Tx and the Rx. More specifically, we define beamforming vector $\mathbf{f}$ at the Tx side and combining vector $\mathbf{w}$ at the Rx side as

\begin{equation}
    \mathbf{f}=\mathbf{a}(N_{\text{Tx}},\phi^{\text{Tx}},\theta^{\text{Tx}}), \quad \mathbf{w}=\mathbf{a}(N_{\text{Rx}},\phi^{\text{Rx}},\theta^{\text{Rx}}).
    \label{BFeq}
\end{equation}

The antenna beamwidth can be adjusted by changing the number of antenna elements $N_{\text{Tx}}$ at the Tx and $N_{\text{Rx}}$ at the Rx. The antenna beamwidth is inversely proportional
to the number of antennas. The more antennas are used, the narrower the
beams will be, which leads to two benefits: The Rx gets a stronger signal
and there is less interference to other non-intended directions. 

We assume that each Tx is capable of having beam codebooks of different sizes. Each beam codebook ($\mathcal{F}$)  consists of a set of beams with the same width. Each vector $\mathbf{f}\in \mathcal{F}$ is of the form \eqref{BFeq} such that the codebook provides complete coverage over all the angular space. The size of the codebook is based on the beamwidth and determines the antenna gain and sidelobe interference caused by each beam. We assume that the main lobes of the different beams of the same codebook are non-overlapping. 
Similarly, we define $\mathcal{W}$ as the combiner codebook at the Rx side.  

Under assumption of capacity achieving codes, UE's achieved rate per second ($R_i$) can be approximated by the link capacity which is defined as:

\begin{equation}
    R_i=B \log_2 (1 + \text{SNR}_i ),
\end{equation}
where $B$ is the system bandwidth and SNR is the signal-to-noise ratio.

\begin{defn}[Trajectory rate]
The trajectory rate ($R_{\text{T}}$) is  the sum of the user's achieved rate along the trajectory with length $M$ as

\begin{equation}
    R_{\text{T}}=\sum_{i \in [M]} R_{i}.
\end{equation}
\end{defn}

\begin{defn}[Path skeleton (PS)]
We define path skeleton as a set of $L$ strongest available paths between the Tx and the Rx. The path skeleton set approximates the spatial channel response of the mmWave networks. Each path $\ell \in [L]$ in path skeleton set with size $L$ can be defined based on its horizontal and vertical AoD ($\phi^{\text{Tx}}$, $\theta^{\text{Tx}}$), horizontal and vertical AoA \mbox{($\phi^{\text{Rx}}$, $\theta^{\text{Rx}}$)} and its gain. An example of a path skeleton set is shown in Fig. \ref{ps}.
\end{defn}

\section{ Beamforming Algorithm}\label{BF}
We adapt a beamforming algorithm that we proposed in \cite{Sara}. In this section, we briefly review our proposed beamforming algorithm and refer readers to \cite{Sara} and \cite{Sara2} for more details and simulation results.
 
We assume that each BS as the Tx has a database of all the path skeleton sets corresponding to user locations in its coverage area.
 Each BS divides its coverage area into non-overlapping grids with a specific ID and pre-defined size. We further assume that the BS represents each grid by one point (the grid centre), and stores one path skeleton for each grid ID in its coverage area. We represent grids each with index $i \in [M]$ where $M$ is the length of the trajectory.
For instance, the path skeleton set in grid $i$ ($\text{PS}_i$) with $L$ paths is
\begin{equation*}
   \text{PS}_i=\{ \phi^{\text{Tx}}_{1},\theta^{\text{Tx}}_{1}, \phi^{\text{Rx}}_{1},\theta^{\text{Rx}}_{1},\beta_1,..., \phi^{\text{Tx}}_{L},\theta^{\text{Tx}}_{L}, \phi^{\text{Rx}}_{L},\theta^{\text{Rx}}_{L},\beta_L\}.
\end{equation*}

In our beamforming method, we assume the location information including the grid ID of the user is available and uses as the input of our algorithm. 

The database assumption causes query cost and maintenance cost. The query cost is defined as the limited budget for each BS to query a path skeleton from its database. The maintenance cost is the cost of building and updating the database. 
In this work, we assume an updated database is available and we will only consider the query cost. We refer the readers to our previous works in  \cite{Sara} and \cite{Sara2} for more details regarding the process of building and updating the database.

Our proposed beamforming method is based on sending pilot signals along the paths in each path skeleton set and not in all directions in order to find the existing non-blocked paths. In other words, the user as the Rx at the grid $i$ asks its path skeleton ($\text{PS}_i$) from its serving BS and measures the signal strength in all $L$ paths in $\text{PS}_i$  and builds $\mathbf{H}_i$ from \eqref{H}.

During the data transmission phase, the beamforming vector, $\mathbf{f}$, and combining vector, $\mathbf{w}$,  are designed as 
\begin{subequations}
	\begin{equation}
	\begin{aligned}
	& \underset{\mathbf{f}, \mathbf{w}}{\text{maximize}}
	&& |\mathbf{w}^H\mathbf{H}\mathbf{f}|^2 
	\end{aligned}
	\end{equation}
	\begin{equation}
	\begin{aligned}
	& \text{subject to}
	& \mathbf{f} \in \mathcal{F}, 
	\end{aligned}
	\end{equation}
	\begin{equation}
	\begin{aligned}
	&&&&&&&&&&\mathbf{w} \in \mathcal{W},
	\end{aligned}
	\end{equation}
	\label{codebook}
\end{subequations}
where $\mathcal{F}$ ($\mathcal{W}$) is the beamforming (combining) codebook contains all the feasible  beamforming (combining) vectors in the form of \eqref{BFeq}.
In an environment with small number of scatters, it is reasonable to adjust beamforming and combining weights to match the array response vector of the strongest path \cite{andrews2016modeling}. That is $\mathbf{f}=\mathbf{a}(\phi_\ell^{\text{Tx}},\theta_\ell^{\text{Tx}})$ and $\mathbf{w}=\mathbf{a}(\phi_\ell^{\text{Rx}},\theta_\ell^{\text{Rx}})$, where $\ell=\argmax_\ell h_\ell$. In summary, finding the beamforming  and combining vectors in the asymptotic regime is based on the Rx feedback to the Tx regarding the index of the path with the maximum received signal strength. For the sake of the notation simplicity, we drop the index $i$ from $\mathbf{f}$ and $\mathbf{w}$.

Afterwards, with designed $\mathbf{f}_\ell$ and $\mathbf{w}_\ell$, the signal to noise ratio (SNR) of the Rx in grid $i$ is

\begin{equation}
    \text{SNR}_i=\sigma^2 |\mathbf{w}_\ell^H \mathbf{H}_i \mathbf{f}_\ell|^2,
\end{equation}
where and $\sigma^2$ is the transmit power divided by the noise power.

Due to the correlation of the spatial channel of mmWave networks, for most mobility models, path skeleton sets are
almost the same over many coherence intervals, essentially over several adjacent locations of a trajectory.
In our proposed method, in order to reduce the unnecessary path skeleton queries and
feedback overhead, the BS continuously tracks the correlation of the path skeletons and
updates them only when a significant change is detected.

We define the reference path skeleton as the path skeleton that BS already reported to the UE in the reference location. We represent the reference location index by $0$. $\mathbf{H}_0$ is the channel matrix at the reference location. When UE changes its location to $i$, the BS sends the pilot signals through the reference path skeleton directions and estimates the channel $\mathbf{H}_i$. In order to validate using the reference path skeleton in the new location we define $d(i,0)$:

\begin{equation}
    d(i,0) = \parallel \mathbf{H}_i-\mathbf{H}_0 \parallel_2,
\end{equation}
where $\lVert \cdot \rVert$ represents the Frobenius norm of the matrix.

We define $\text{T}_\text{D}$ as the distance threshold that is a small positive number. The condition $d(i,0)\leq \text{T}_\text{D}$ means the validity of using reference path skeleton in new location $i$. Otherwise,  $d(i,0) > \text{T}_\text{D}$ translates to a significant change in dominant paths, and the BS needs to query a new path skeleton and inform the UE.  Now, the new
path skeleton is considered the reference path skeleton and the BS tracks the validity
of this new reference path skeleton for beam-searching and channel estimation over time. Note that path skeleton sets may not be changed with every new obstacle. A significant change in the environment may change the path skeleton sets.  

There is a trade-off between the value of $\text{T}_\text{D}$ and the rate and the query budget. We define the query budged as the number of times that BS queries a path skeleton from the database.  In other words, a smaller
value of $\text{T}_\text{D}$  improves the rate performance but with the expense of higher query cost. A larger value of $\text{T}_\text{D}$  leads to lower network overhead, however, may cause sub-optimal
selection of the beamforming vector and combining vector along the UE trajectory.
The optimal value of $\text{T}_\text{D}$ for a specific mobility model (e.g. pedestrian) is the solution of 
following optimization problem:
\begin{subequations}
	\begin{equation}
	\begin{aligned}
	& \underset{\text{T}_\text{D}>0}{\text{${\text{T}_\text{D}^\star}$ = $\mathrm{argmax}$}}
	\sum_{i \in [M]}\mathbb{E}[R_{i}]
	\end{aligned}
	\end{equation}
	\begin{equation}
	\begin{aligned}
	& \text{subject to}
	&& \text{Pr} \{U>U_{\max}\}\leq \delta,
	\end{aligned}
	\end{equation}
	\label{T}
\end{subequations}
where the expectation is with respect to the channel randomness (small scale fading and blockage), $M$ is the length of the trajectory and $U$ is the number of times that path skeletons are renewed (query cost). $U_\text{max}$ is the maximum query cost. $\delta$ is a small given parameter. In order to solve the \eqref{T}, we apply a well-known golden-section search method \cite{Chong}. Notice that the process of finding $\text{T}^\star_\text{D}$ is run offline and doesn't add overhead to the real-time system.



\section{Beamforming under Errors in Location Information}\label{BF with e}
In this section, first, we define the location information model. 
Then we present the robust beamforming in the presence of location error.


\begin{figure}[t]
	\centering

		\includegraphics[width=0.26\textwidth]{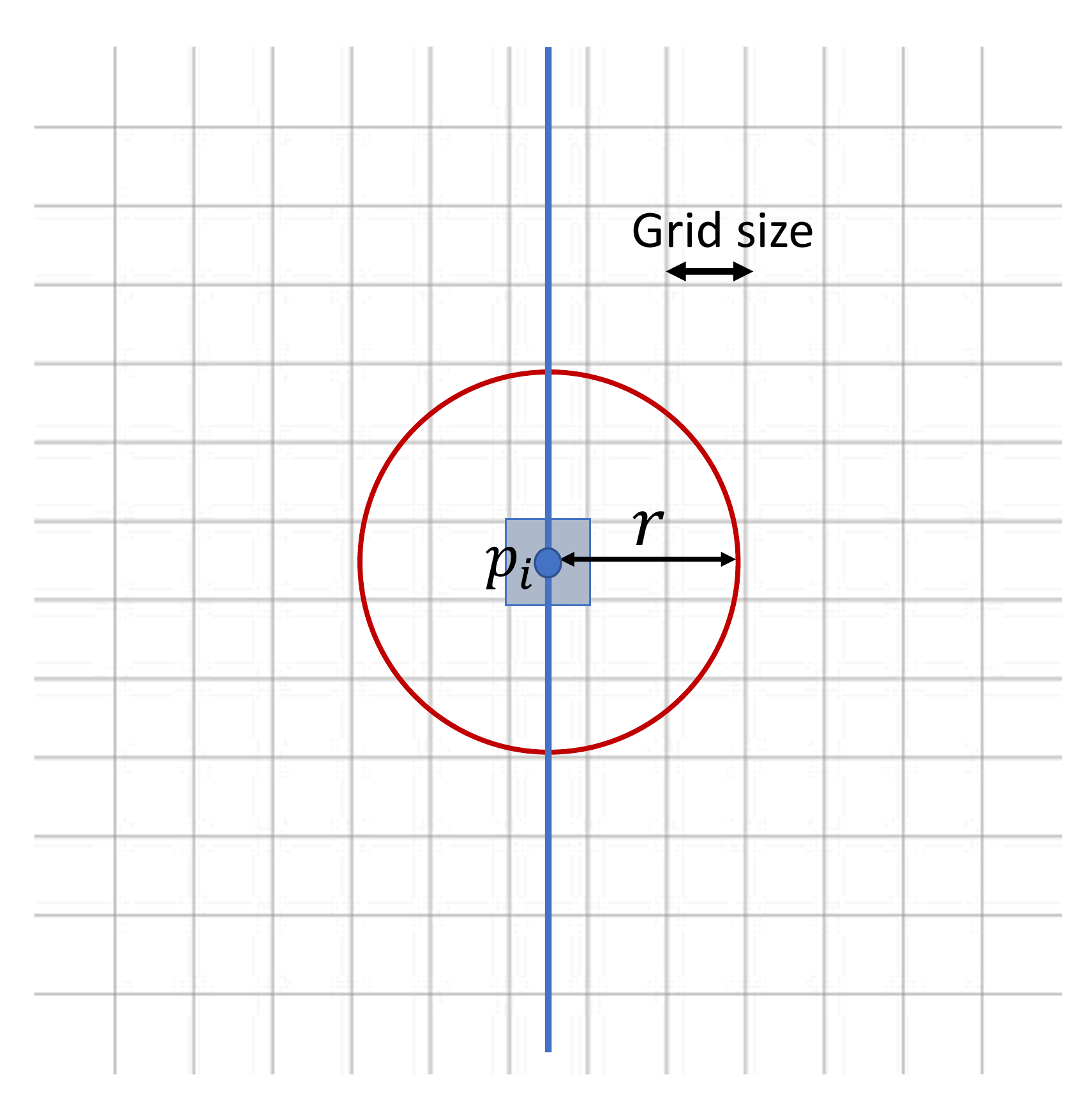}
			\caption{Localization error model. Blue line shows a potential trajectory. The area is divided into grids with equal size. The uncertainty disk $E(r) $ is the red circle. Grid $i$ is marked by blue. Estimated user location, $\hat{p}_i$, could be anywhere inside the red circle. Depending on the uncertainty radius, user location can be mistakenly estimated to be in another grid.  }
			\label{r}
\end{figure}
\subsection{Localization Error Model}

The input of our proposed beamforming method in the previous section is the location of the user along its trajectory.
In a realistic setting where both BS and user separately obtain location information via some localization processes, noisy positional information occurs \cite{maschietti2017robust}\footnote{Details regarding the performance of the current positioning methods in cellular networks are reported in \cite [Table III]{del2017survey}. }. 
The noisy position $\hat{p}$ available at the BS is modeled as:

\begin{equation}
    \hat{p}_i=p_i+e_i,
\end{equation}
where $p$ is actual location of the user.
We apply a uniform bounded error model for location information. In particular. we assume all the estimated locations (as the output of different location services) lie somewhere inside an uncertainty disk, $E(r)$, centred in the actual origin of a  grid ($\mathbf{p}_i$, $i \in \lbrack M \rbrack$)  with radius $r$. Fig. \ref{r} illustrates $E(r)$ in one of the grid points along a trajectory.
We model the random localization error, $\mathbf{e}$, as uniformly distributed in $E(r)$.

Different localization algorithms correspond to different uncertainty radius, $r$. A bigger r may translate into a potentially cheaper localization algorithm. However, when the location is mistakenly set to grid $j$ instead of $i$ (the distance between $i$ and $j$ is a function of $r$), the user will use the path skeleton of grid j for its beamforming design, leading to a potential performance drop.  

\subsection{Robust Beamforming}

As mentioned in the previous section, we approximate each grid with a pre-defined size as a point and store one path skeleton for each grid in the database. The user's grid IDs along the trajectory are the input of our beamforming method. Hence, the performance of our proposed beamforming method depends on the location information inputs. However, due to the spatial correlation of mmWave channels, path skeletons might be similar in adjacent grids. Now the question is how accurate location information we need as an input to the beamforming algorithm.

First, we consider an idealized case where the perfect location information is available at the BS. In this case, during the beam searching phase, the BS extracts the path skeleton of the grid $i$ and starts beam searching over the $\text{PS}_i$ set. The optimal beam directions can be found as 

\begin{equation}
   \underset{\mathcal{D}_{\text{Tx}} \subset PS_i,\mathcal{D}_{\text{Rx}} \subset PS_i }{(\mathcal{D}^\star_{\text{Tx}},\mathcal{D}^\star_{\text{Rx}})= \mathrm{argmax} \quad R(\mathcal{D}_{\text{Tx}},\mathcal{D}_{\text{Rx}}, p_i)
   },
\end{equation}
where $\mathcal{D}_{\text{Tx}}$ contains the horizontal and vertical AoA and $\mathcal{D}_{\text{Rx}}$  contains the horizontal and vertical AoD.

Now, we consider the noisy location information.
In this case, we focus on the maximum location error that our beamforming algorithm can tolerate. In other words, we need to find the maximum $r$ that 
\begin{subequations}
	\begin{equation}
	\begin{aligned}
	& \underset{r\in \Gamma }{r^\star=\mathrm{argmax}} {\quad[\min_i (\mathbb{E}[R_{i}(r)] )]}
	\end{aligned}
	\end{equation}
	\begin{equation}
	\begin{aligned}
	& \text{subject to}
	&& \text{Pr} \{U>U_{\max}\}\leq \delta
	\end{aligned}
		\label{eb}
	\end{equation}
		\begin{equation}
	\begin{aligned}
	 &&& R_i(r) \geq R_{th},
	\end{aligned}
	\label{ec}
	\end{equation}
	\label{e}
\end{subequations}where $\Gamma$ is a set of predefined $r$ that correspond to a set of localization algorithms each with certain accuracy. The constraint \eqref{eb} refers to the limited query budget (path skeleton updates) of the database and constraint \eqref{ec} guarantees that the achieved rate of the user in any grid $i \in [M] $ of the trajectory is larger than a pre-defined threshold ($R_{th}$) and ensures a reliable connection along the trajectory. We solve \eqref{e} numerically by comparing the solutions for all $r\in\Gamma$.


\section{Simulation Results}\label{simulations}
In this section, we present the performance evaluation of our proposed approach. First, we introduce our simulation setup. Then, we present the location input model and finally, we evaluate the performance of our proposed beamforming method with different degrees of location information precision.



\subsection{Simulation Setup}
\begin{figure}[t]
	\centering
		\includegraphics[width=0.45\textwidth]{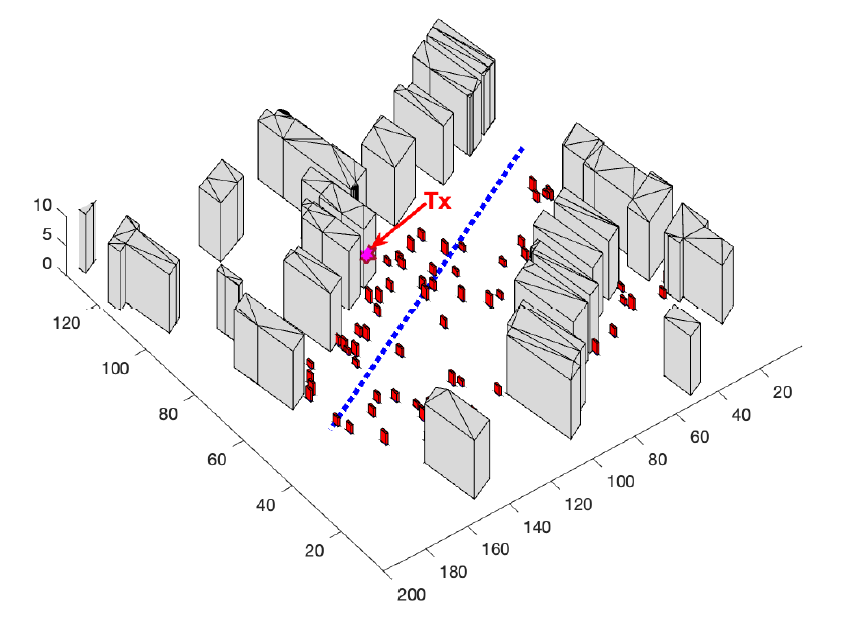}
	\caption{Simulation environment. Blue dottd line illustrates the trajectory. Pink hexagram shows the location of the Tx.  }
	\label{scenario}
\end{figure}

\begin{table}[bp]
	\begin{center}
		\caption{Simulation parameters. }
		\label{table1}
		\begin{tabular}{|c|c|} 
			\hline
			\textbf{Parameters} & \textbf{Values in Simulations} \\
			\hline
			\hline 
		     transmit power & 30 dBm\\
			Thermal noise power	&$\sigma^2$=-174 dBm/Hz\\
			Signal bandwidth & $B$=100 MHz \\
			Operation frequency&28 GHz \\
			LoS path loss exponent	 & $1.9$ \\
			NLoS path loss exponent & $4.5$ \\
			BS height	 & 6 m\\
			Brick penetration loss \cite{penetration} & 28.3 dB\\
			Glass penetration loss \cite{penetration} & 3.9 dB\\
			\hline 
		\end{tabular}
	\end{center}
\end{table}
We consider an urban environment and apply a ray-tracing tool \cite{simic2017demo} in order to obtain the existing paths between the Tx and the Rx.
To ensure
high angular resolutions, we measure the AoAs and the AoDs with step sizes of 0.1 degrees.
The simulation environment is Shown in Fig. \ref{scenario}. We consider the buildings as permanent obstacles and randomly assign brick or glass materials to them. We also add some temporary random obstacles in the street with heights $1.5$ m as the human bodies and some random obstacles with widths $4$ m and heights \mbox{$1$ m} and $3$ m to model various vehicles with different heights. Temporary obstacles are located uniformly on the streets with a density of $9 \times 10^{-3}$ per $m^2$. The material loss of the temporary obstacles is chosen randomly in each realization of the channel. Red cubes in  \mbox{Fig. \ref{scenario}} show one realization of temporary obstacles.

We place the UE's trajectory with length $150$ m and apply the pedestrian mobility model with a speed of $5$ km/h \footnote{We will consider other mobility models with different speeds in our future works.}. In all simulations, we fix the trajectory. We consider the grid size equal to $3$ m which means we have $50$ grids with index $i \in [50] $ along the trajectory. We also place the Tx on the wall of the building (see Fig. \ref{scenario}) around the middle of the street. We consider that all the locations of the trajectory are in the coverage area of the Tx.
We consider two sets of antenna elements: ($N_{\text{BS}}=64$, $N_{\text{UE}}=16$) represents a narrow beamwidth antenna regime (as the first scenario)  and \mbox{($N_{\text{BS}}=32$,} $N_{\text{UE}}=8$) represents a wide antenna beamwidth regime (as the second scenario).
We also set \mbox{$R_{\text{th}}=200$ Mbps} \mbox{in \eqref{e}}.  The general simulation parameters are listed in \mbox{Table I}.


 \begin{figure}
	\centering
	\begin{subfigure}{0.5\textwidth}
		{\footnotesize 
%
%

\pgfplotsset{major grid style = {dotted, gray}}
\pgfplotsset{minor grid style = {dotted, gray}}
\begin{tikzpicture}

\begin{axis}[%
width=0.8\columnwidth,
height=0.4\columnwidth,
at={(0\columnwidth,0\columnwidth)},
scale only axis,
xmin=0,
xmax=52,
xlabel style={font=\color{white!15!black}},
xlabel={Grid point index},
ymin=0,
ymax=1,
ylabel style={font=\color{white!15!black}},
ylabel={$R$ (Gbps)},
axis background/.style={fill=white},
xmajorgrids,
ymajorgrids,
legend style={at={(0.02,0.459)}, anchor=south west, legend cell align=left, align=left, draw=white!15!black}
]
\addplot [color=blue, line width=1.0pt]
  table[row sep=crcr]{%
1	0.340783624437366\\
2	0.331072502301245\\
3	0.316427629696591\\
4	0.32756225522084\\
5	0.346933787923728\\
6	0.346213376401011\\
7	0.342290520485792\\
8	0.348370491483491\\
9	0.352622412064282\\
10	0.327530392547277\\
11	0.352766716801662\\
12	0.390127281179496\\
13	0.356957720034887\\
14	0.331977657012441\\
15	0.232995527039028\\
16	0.264974337347207\\
17	0.397677285051409\\
18	0.458306725009769\\
19	0.507597249217899\\
20	0.530567039102473\\
21	0.469876620392797\\
22	0.591578841515452\\
23	0.57379928782558\\
24	0.570471930190204\\
25	0.527015797819828\\
26	0.697614488061325\\
27	0.65977602115719\\
28	0.75665877174756\\
29	0.551212431980645\\
30	0.647465015559689\\
31	0.682614174448845\\
32	0.495907237736945\\
33	0.664433104712077\\
34	0.52873024317139\\
35	0.743823956414427\\
36	0.408136589274623\\
37	0.383766725172509\\
38	0.508396891758369\\
39	0.385291287961316\\
40	0.377611923641357\\
41	0.375371164273831\\
42	0.373161176692941\\
43	0.372514742695266\\
44	0.371125468776882\\
45	0.36827053753404\\
46	0.368583313884427\\
47	0.360704832278331\\
48	0.367654103919523\\
49	0.360188095109458\\
50	0.364084368165535\\
51	0.357335458123467\\
};
\addlegendentry{$r=0$}

\addplot [color=red, line width=1.0pt]
  table[row sep=crcr]{%
1	0.299067702136122\\
2	0.330153953307075\\
3	0.3289426297907\\
4	0.346053653923477\\
5	0.344915385867886\\
6	0.339624015494589\\
7	0.380025104337742\\
8	0.35403139709753\\
9	0.335461534741505\\
10	0.335143552255898\\
11	0.353489671795871\\
12	0.383226224824553\\
13	0.355266952897399\\
14	0.357582822314776\\
15	0.309253094032275\\
16	0.370212094856336\\
17	0.3760088614226\\
18	0.440769887549163\\
19	0.457220813152167\\
20	0.441408503302276\\
21	0.508290662046999\\
22	0.585511547108953\\
23	0.53032107767387\\
24	0.602002398816824\\
25	0.561722710386392\\
26	0.842085351930724\\
27	0.671589256251941\\
28	0.783417620573977\\
29	0.677764592337894\\
30	0.666311429406735\\
31	0.738824666705167\\
32	0.649930830188319\\
33	0.735176461983068\\
34	0.635467566863362\\
35	0.845358197622215\\
36	0.401098009250115\\
37	0.393306204925085\\
38	0.510902331576232\\
39	0.391626110322013\\
40	0.372593469191124\\
41	0.373258001582209\\
42	0.364376226010886\\
43	0.337593285870867\\
44	0.346993371347227\\
45	0.316943024561808\\
46	0.264165299698568\\
47	0.301828700018551\\
48	0.270047027865021\\
49	0.25423591280766\\
50	0.255916158167762\\
51	0.236658859614816\\
};
\addlegendentry{$r=10$ m}

\addplot [color=green,line width=1.0pt]
  table[row sep=crcr]{%
1	0.283138953763267\\
2	0.345125090941553\\
3	0.311902848553221\\
4	0.33399355123987\\
5	0.345255666823887\\
6	0.363609307051315\\
7	0.356206093720438\\
8	0.351788272514037\\
9	0.362258238084607\\
10	0.343545981123182\\
11	0.357480101375841\\
12	0.349771003235203\\
13	0.344404408905371\\
14	0.34145191496934\\
15	0.319245697188073\\
16	0.379219720023709\\
17	0.439046790731686\\
18	0.40771563092455\\
19	0.491306314558174\\
20	0.515762177960339\\
21	0.481468194322865\\
22	0.534206503508533\\
23	0.568916591503733\\
24	0.611158927226452\\
25	0.5953691669685\\
26	0.796910739350197\\
27	0.613452667902497\\
28	0.754897158739381\\
29	0.633005612476407\\
30	0.703198433885581\\
31	0.762858982485823\\
32	0.702877779546161\\
33	0.828283483838417\\
34	0.585930434047551\\
35	0.807309019776735\\
36	0.406814708747446\\
37	0.374719945880499\\
38	0.496574767672785\\
39	0.363991877826976\\
40	0.359723563382254\\
41	0.339430327287508\\
42	0.336004158299715\\
43	0.316384355427892\\
44	0.300364869387368\\
45	0.284416071021966\\
46	0.2951276347806\\
47	0.276367104734022\\
48	0.252066346286497\\
49	0.247472258620466\\
50	0.20264410308611\\
51	0.199173138308887\\
};
\addlegendentry{$r=11$ m}

\addplot [color=black,line width=1.0pt]
  table[row sep=crcr]{%
1	0.266273508018848\\
2	0.344656499671418\\
3	0.326218479786673\\
4	0.323641974893309\\
5	0.360641749669025\\
6	0.3573106298912\\
7	0.385725366226215\\
8	0.362880743085751\\
9	0.35674735984343\\
10	0.337916519943994\\
11	0.364690813008429\\
12	0.32509928076294\\
13	0.395910631758498\\
14	0.390288044391787\\
15	0.299946641361048\\
16	0.40205960599184\\
17	0.419244908485975\\
18	0.415626949058486\\
19	0.441200585483488\\
20	0.502465718119437\\
21	0.487437145902269\\
22	0.528475293636876\\
23	0.599000195881054\\
24	0.588820827427715\\
25	0.554879296738539\\
26	0.732905925449786\\
27	0.628797702490609\\
28	0.834961859581475\\
29	0.691142411685433\\
30	0.72201835187487\\
31	0.771564088641567\\
32	0.649946403787963\\
33	0.781757531809501\\
34	0.628610541737378\\
35	0.780412940836795\\
36	0.390384339489316\\
37	0.39880542078752\\
38	0.515449287238478\\
39	0.360038709692885\\
40	0.325560155241051\\
41	0.284823340482331\\
42	0.303564459392011\\
43	0.237061743608965\\
44	0.249061442106684\\
45	0.222137880040475\\
46	0.194134449164616\\
47	0.188527455898781\\
48	0.151867685849396\\
49	0.181319357794263\\
50	0.140897506661175\\
51	0.108149588687604\\
};
\addlegendentry{$r=12$ m}

\addplot [color=purple, dashed, line width=1.0pt]
  table[row sep=crcr]{%
1	0.2\\
2	0.2\\
3	0.2\\
4	0.2\\
5	0.2\\
6	0.2\\
7	0.2\\
8	0.2\\
9	0.2\\
10	0.2\\
11	0.2\\
12	0.2\\
13	0.2\\
14	0.2\\
15	0.2\\
16	0.2\\
17	0.2\\
18	0.2\\
19	0.2\\
20	0.2\\
21	0.2\\
22	0.2\\
23	0.2\\
24	0.2\\
25	0.2\\
26	0.2\\
27	0.2\\
28	0.2\\
29	0.2\\
30	0.2\\
31	0.2\\
32	0.2\\
33	0.2\\
34	0.2\\
35	0.2\\
36	0.2\\
37	0.2\\
38	0.2\\
39	0.2\\
40	0.2\\
41	0.2\\
42	0.2\\
43	0.2\\
44	0.2\\
45	0.2\\
46	0.2\\
47	0.2\\
48	0.2\\
49	0.2\\
50	0.2\\
51	0.2\\
};
\addlegendentry{Threshold}

\end{axis}
\end{tikzpicture}%
}
		
		\caption{}
		\label{22a}
	\end{subfigure}
	\begin{subfigure}[t]{0.5\textwidth}
		{\footnotesize \input{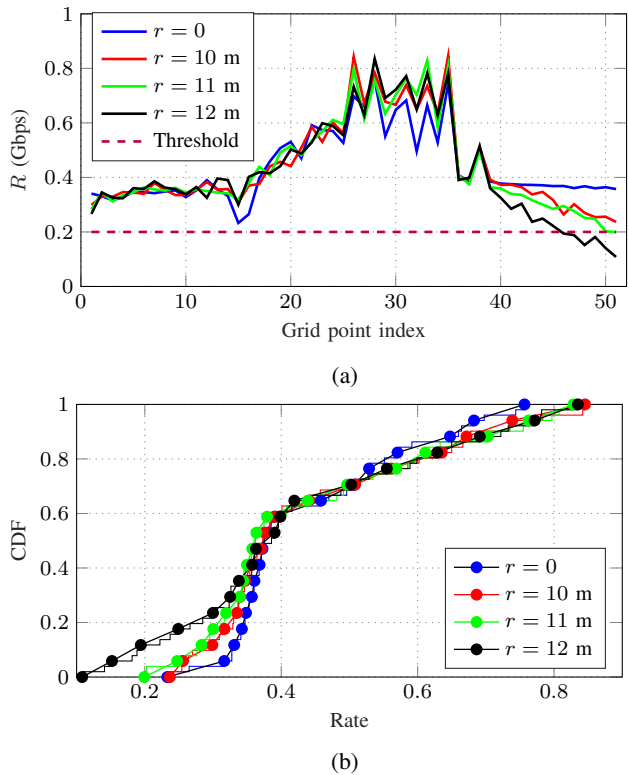}}
		
		\caption{}
		\label{22b}
	\end{subfigure}
		
	\caption{(a) Achieved rate per grid (b) distribution of the rate for the first scenario. }
	\label{2d_adj}
\end{figure}

\subsection{Location Error and Rate Trade-off}
\begin{figure}
	\centering
	\begin{subfigure}[t]{0.5\textwidth}
		{\footnotesize 
%
%

%
\pgfplotsset{major grid style = {dotted, gray}}
\pgfplotsset{minor grid style = {dotted, gray}}
\begin{tikzpicture}

\begin{axis}[%
width=0.80\columnwidth,
height=0.4\columnwidth,
at={(0\columnwidth,0\columnwidth)},
scale only axis,
xmin=0,
xmax=52,
xlabel style={font=\color{white!15!black}},
xlabel={Grid point index},
ymin=0,
ymax=0.75,
axis background/.style={fill=white},
xmajorgrids,
ymajorgrids,
ylabel style={font=\color{white!15!black}},
ylabel={$R$},
axis background/.style={fill=white},
legend style={at={(0.02,0.459)}, anchor=south west, legend cell align=left, align=left, draw=white!15!black}
]
\addplot [color=blue,line width=1.0pt]
  table[row sep=crcr]{%
1	0.271445407087475\\
2	0.262421901037201\\
3	0.257643574920793\\
4	0.253110756975047\\
5	0.276272833757342\\
6	0.260406162800235\\
7	0.267690635996391\\
8	0.279814277337483\\
9	0.277592718425907\\
10	0.263913980816143\\
11	0.268804022760566\\
12	0.265166848051465\\
13	0.256934932744497\\
14	0.253409795150207\\
15	0.242431617731955\\
16	0.230525228426914\\
17	0.233550777490345\\
18	0.266962399071395\\
19	0.273688284806247\\
20	0.341759284997487\\
21	0.364536154227106\\
22	0.378073356029474\\
23	0.343384030884915\\
24	0.404160100339993\\
25	0.35777564509289\\
26	0.411206488928126\\
27	0.443797566729005\\
28	0.508410210016896\\
29	0.459828084695223\\
30	0.419610231455561\\
31	0.406687853295393\\
32	0.366910969968372\\
33	0.411217434205158\\
34	0.362047213704447\\
35	0.52530702891307\\
36	0.329485323721001\\
37	0.303281774283578\\
38	0.334429771719677\\
39	0.311365283269493\\
40	0.31209903451233\\
41	0.31153601922024\\
42	0.307310392054428\\
43	0.305936158532162\\
44	0.302648161142169\\
45	0.297364195075823\\
46	0.300886004777806\\
47	0.296019564956269\\
48	0.300540223423806\\
49	0.295028468971048\\
50	0.295211629853381\\
51	0.298157643044528\\
};
\addlegendentry{$r=0$}

\addplot [color=red,line width=1.0pt]
  table[row sep=crcr]{%
1	0.220032001538709\\
2	0.273160579840927\\
3	0.260187666302324\\
4	0.266716585478055\\
5	0.263557787029599\\
6	0.278355855588736\\
7	0.268003030919701\\
8	0.271297442354639\\
9	0.273097912168176\\
10	0.260864989428981\\
11	0.270846009720186\\
12	0.267622320932823\\
13	0.263101627094209\\
14	0.266915663375568\\
15	0.265760483056245\\
16	0.276383704723115\\
17	0.30449217687171\\
18	0.341583549104257\\
19	0.326564187691689\\
20	0.320915919695609\\
21	0.338750943352539\\
22	0.367262551354387\\
23	0.400136148306872\\
24	0.385051761802529\\
25	0.379527884592674\\
26	0.450098713541201\\
27	0.421203290035069\\
28	0.497060329624479\\
29	0.396072307051146\\
30	0.430556614197291\\
31	0.466795549973543\\
32	0.437733245727248\\
33	0.499848787664898\\
34	0.474295480191274\\
35	0.507801695050374\\
36	0.324581005708584\\
37	0.321917542584844\\
38	0.361738992880854\\
39	0.30382442201885\\
40	0.286128816866951\\
41	0.278530435827563\\
42	0.265947681839639\\
43	0.272383450439361\\
44	0.25767074888212\\
45	0.282465284525034\\
46	0.249840894607782\\
47	0.239563645594081\\
48	0.237202889206409\\
49	0.230062518086247\\
50	0.224461211352836\\
51	0.234708616375944\\
};
\addlegendentry{$r=19$ m}

\addplot [color=green,line width=1.0pt]
  table[row sep=crcr]{%
1	0.19244796849672\\
2	0.262982910383524\\
3	0.261050423309085\\
4	0.259522320823959\\
5	0.26697775586521\\
6	0.260175263471025\\
7	0.272641775806334\\
8	0.279872265402294\\
9	0.279038727236996\\
10	0.26441653517911\\
11	0.276183123150541\\
12	0.28421451044056\\
13	0.272595414978696\\
14	0.272092021290097\\
15	0.26697077094369\\
16	0.288214943201559\\
17	0.325864240940728\\
18	0.308130366452149\\
19	0.30138891829307\\
20	0.339164684549353\\
21	0.332581804093762\\
22	0.379238872484244\\
23	0.378393752312687\\
24	0.421815256114589\\
25	0.403897971139657\\
26	0.525621134865079\\
27	0.468319743745321\\
28	0.460928446620264\\
29	0.465070582932721\\
30	0.403513787705518\\
31	0.519491911771706\\
32	0.499456339172484\\
33	0.533520418814496\\
34	0.415902270477974\\
35	0.479515405424864\\
36	0.322078322746682\\
37	0.289415164270258\\
38	0.339258408007488\\
39	0.290000745758723\\
40	0.262759898776907\\
41	0.250437660898337\\
42	0.235338511768221\\
43	0.232539166323569\\
44	0.240232731197155\\
45	0.220388645843091\\
46	0.236580469852724\\
47	0.204633545377982\\
48	0.219278853119707\\
49	0.193842652398565\\
50	0.213240121110938\\
51	0.185787673697959\\
};
\addlegendentry{$r=20$ m}

\addplot [color=black,line width=1.0pt]
  table[row sep=crcr]{%
1	0.204152020748637\\
2	0.26468513551849\\
3	0.25567623219622\\
4	0.261227529559356\\
5	0.260405702166224\\
6	0.247295382277321\\
7	0.265192703524556\\
8	0.267193577492789\\
9	0.271536337155246\\
10	0.264220523462783\\
11	0.255511637967058\\
12	0.274086485348863\\
13	0.275845208886375\\
14	0.283312914700642\\
15	0.265257074936477\\
16	0.276472038280553\\
17	0.315554628793067\\
18	0.289444714010881\\
19	0.323963545775173\\
20	0.352630147445144\\
21	0.359293583436458\\
22	0.373539138151245\\
23	0.393219452141533\\
24	0.412112278423806\\
25	0.369223664902222\\
26	0.496756966527946\\
27	0.44002256847409\\
28	0.539176762144381\\
29	0.415839779773328\\
30	0.4659512359452\\
31	0.503779394512819\\
32	0.434664329744001\\
33	0.504111447114872\\
34	0.386995848633907\\
35	0.486492456001605\\
36	0.301188509187359\\
37	0.306408209630151\\
38	0.324360476326488\\
39	0.267772116260964\\
40	0.234355625817736\\
41	0.251542715275418\\
42	0.220525412115659\\
43	0.201029903122259\\
44	0.200003622758514\\
45	0.223224737202186\\
46	0.199082041315584\\
47	0.214919023862642\\
48	0.22592992863779\\
49	0.198245137477503\\
50	0.159943849774597\\
51	0.208326375144472\\
};
\addlegendentry{$r=21$ m}

\addplot [color=purple, dashed, line width=1.0pt]
  table[row sep=crcr]{%
1	0.2\\
2	0.2\\
3	0.2\\
4	0.2\\
5	0.2\\
6	0.2\\
7	0.2\\
8	0.2\\
9	0.2\\
10	0.2\\
11	0.2\\
12	0.2\\
13	0.2\\
14	0.2\\
15	0.2\\
16	0.2\\
17	0.2\\
18	0.2\\
19	0.2\\
20	0.2\\
21	0.2\\
22	0.2\\
23	0.2\\
24	0.2\\
25	0.2\\
26	0.2\\
27	0.2\\
28	0.2\\
29	0.2\\
30	0.2\\
31	0.2\\
32	0.2\\
33	0.2\\
34	0.2\\
35	0.2\\
36	0.2\\
37	0.2\\
38	0.2\\
39	0.2\\
40	0.2\\
41	0.2\\
42	0.2\\
43	0.2\\
44	0.2\\
45	0.2\\
46	0.2\\
47	0.2\\
48	0.2\\
49	0.2\\
50	0.2\\
51	0.2\\
};
\addlegendentry{Threshold}

\end{axis}
\end{tikzpicture}%
}
		
		\caption{}
		\label{}
	\end{subfigure}
	\begin{subfigure}[t]{0.5\textwidth}
		{\footnotesize \input{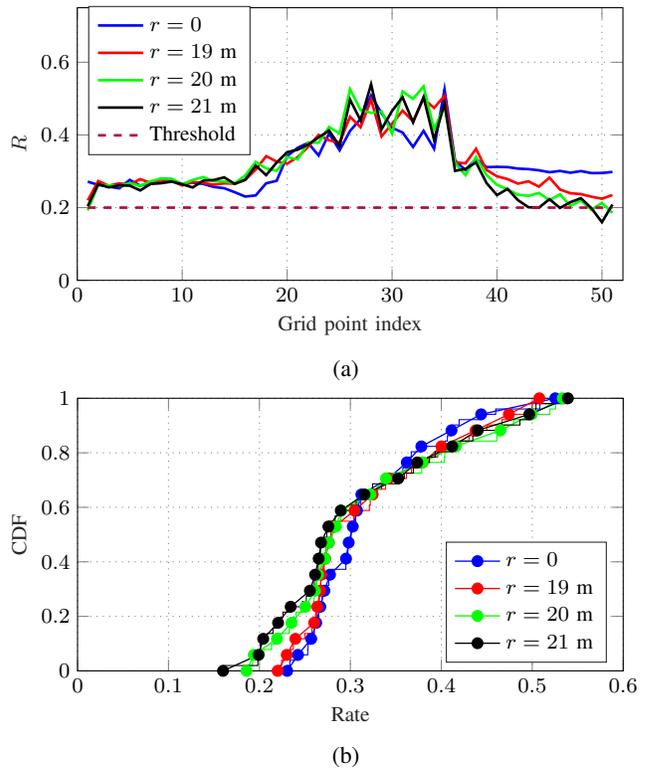}}
		
		\caption{}
		\label{}
	\end{subfigure}
	\caption{(a) Achieved rate per grid (b) distribution of the rate for the second scenario. }
	\label{6d_adj}
\end{figure}

Fig. \ref{22a} shows the achieved rate of the UE in each grid of the trajectory for different values of the $r$ for the first scenario ($N_{\text{BS}}=64$, $N_{\text{UE}}=16$). We only represent the maximum values of the $r$ that meet the rate threshold. We set the maximum number of path skeleton updates ($U_{\max}$) equals $20$. 
As it is shown in this figure, our proposed beamforming method can tolerate the location information error up to \mbox{$11$ m} while meeting the rate threshold and the maximum updating path skeletons. The distribution of the rates in each grid is almost the same with increasing $r$ from $0$ to $11$ m in this scenario. The reason is that our algorithm by tracking the correlation of the path skeletons updates the path skeletons and prevents the sudden rate drops due to the location errors However, by increasing the $r$ ($r\geq 12$ m) we can observe the achieved rate reduction, especially at the end of the trajectory which indicates the trade-off between the achieved rate and the amount of location error. We can observe the same trend in the second scenario in Fig. \ref{6d_adj}. In this case, for $r\geq 20$ m the achieved rate drops below the rate threshold. 


\begin{table}
	\begin{center}
		\caption{Number of path skeleton updates, $U$,  for different value of $r$ in the first scenario (narrow beamwidth) with $U_{\max}=20$ and the second scenario (wide beamwidth) with $U_{\max}=15$. }
		\label{table1}
	\begin{tabular}{|c|c|c|} 
			\hline
 $r$ (m) & $U$ (First scenario) & $U$ (Second scenario)  \\
			\hline
			\hline 
     0  & 14 & 7\\
		      5 & 15 & 8\\
		     10 & 20 &10\\
		     20 & - & 15\\  \hline
\end{tabular}

	\end{center}
\end{table}

\subsection{Location Error and Beamforming Overhead Trade-off}

We define the beamforming overheat as the number of times that the BS needs to update the path skeletons along the trajectory. In the first scenario, we set the maximum number of path skeleton updates ($U_{\max}$) equals $20$ and in the second scenario, we set $U_{\max}$ equals $15$. As it is shown in \mbox{Table II,}  with increasing the location error, the number of the path skeleton updates will be increased too. It means that the path skeletons need to be updated more frequently due to the location information uncertainties. In other words, there is a trade-off between the location error and the beamforming overhead.

\subsection{Location error and Beamwidth trade-off}
Fig.\ref{6d_adj} illustrates the value and the distribution of the of the achieved rate for the second scenario ($N_{\text{BS}}=32$, $N_{\text{UE}}=8$). In this scenario, we set the $U_{\max}=15$. Same as the first scenario, the value and the distribution of the achieved rate are almost the same for $r$ up to $20$ m. Our proposed algorithm by tracking the correlation of the path skeletons and updating when significant changes occur in the environment can tolerate the location errors while meeting the rate threshold with a small number of path skeleton updates along the trajectory. 

In comparison to the first scenario, we can see that wider beams can tolerate more location errors. The reason is that a larger beamwidth leads to a larger coverage of the geographical area and more tolerance of the location uncertainties. However, the larger beamwidth results in a lower radiated power too. Thus, there is a trade-off between the antenna beamwidth and the achieved rate and antenna beamwidth and the localization error tolerance.

\subsection{Benchmarks}
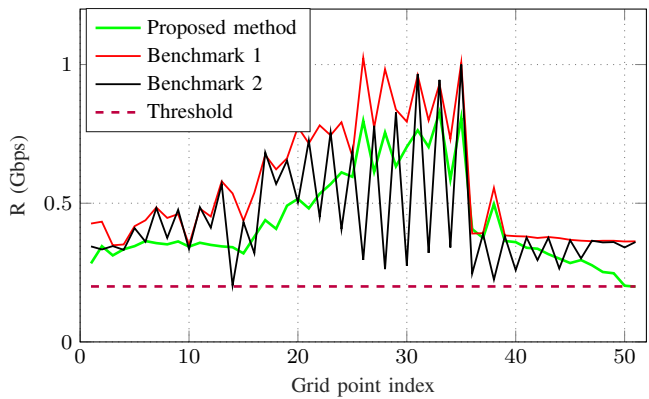
\begin{figure}[t]
	\centering
		{\footnotesize 
%
%

\pgfplotsset{major grid style = {dotted, gray}}
\pgfplotsset{minor grid style = {dotted, gray}}
\begin{tikzpicture}

\begin{axis}[%
width=0.85\columnwidth,
height=0.5\columnwidth,
at={(0\columnwidth,0\columnwidth)},
scale only axis,
xmin=0,
xmax=52,
xlabel style={font=\color{white!15!black}},
xlabel={Grid point index},
ymin=0,
ymax=1.2,
ylabel style={font=\color{white!15!black}},
ylabel={R (Gbps)},
axis background/.style={fill=white},
xmajorgrids,
ymajorgrids,
legend style={at={(0.01,0.63)}, anchor=south west, legend cell align=left, align=left, draw=white!13!black}
]
\addplot [color=green,line width=1.0pt]
  table[row sep=crcr]{%
1	0.283138953763267\\
2	0.345125090941553\\
3	0.311902848553221\\
4	0.33399355123987\\
5	0.345255666823887\\
6	0.363609307051315\\
7	0.356206093720438\\
8	0.351788272514037\\
9	0.362258238084607\\
10	0.343545981123182\\
11	0.357480101375841\\
12	0.349771003235203\\
13	0.344404408905371\\
14	0.34145191496934\\
15	0.319245697188073\\
16	0.379219720023709\\
17	0.439046790731686\\
18	0.40771563092455\\
19	0.491306314558174\\
20	0.515762177960339\\
21	0.481468194322865\\
22	0.534206503508533\\
23	0.568916591503733\\
24	0.611158927226452\\
25	0.5953691669685\\
26	0.796910739350197\\
27	0.613452667902497\\
28	0.754897158739381\\
29	0.633005612476407\\
30	0.703198433885581\\
31	0.762858982485823\\
32	0.702877779546161\\
33	0.828283483838417\\
34	0.585930434047551\\
35	0.807309019776735\\
36	0.406814708747446\\
37	0.374719945880499\\
38	0.496574767672785\\
39	0.363991877826976\\
40	0.359723563382254\\
41	0.339430327287508\\
42	0.336004158299715\\
43	0.316384355427892\\
44	0.300364869387368\\
45	0.284416071021966\\
46	0.2951276347806\\
47	0.276367104734022\\
48	0.252066346286497\\
49	0.247472258620466\\
50	0.20264410308611\\
51	0.199173138308887\\
};
\addlegendentry{Proposed method}
\addplot [color=red,line width=0.7pt]
  table[row sep=crcr]{%
1	0.426702654516927\\
2	0.433514091630429\\
3	0.348350936779632\\
4	0.351340403615165\\
5	0.417178624006479\\
6	0.438511217995616\\
7	0.484664716473464\\
8	0.447000558568538\\
9	0.461027502844772\\
10	0.350076370429068\\
11	0.480711615403344\\
12	0.452141870272161\\
13	0.578867350762694\\
14	0.535415959627741\\
15	0.437175385676227\\
16	0.53779021842324\\
17	0.67389635774194\\
18	0.622223749402467\\
19	0.661963932648416\\
20	0.774461242888882\\
21	0.71575135948266\\
22	0.780700227343626\\
23	0.74499545676468\\
24	0.791864600902026\\
25	0.669153256450843\\
26	1.02407135169999\\
27	0.772506739424551\\
28	0.98323492914821\\
29	0.837667496652907\\
30	0.793981609613266\\
31	0.961412194498266\\
32	0.798866481248945\\
33	0.930241991138333\\
34	0.731951238003686\\
35	1.00680477250537\\
36	0.390740588268054\\
37	0.392107591648454\\
38	0.555435969478192\\
39	0.383874095712236\\
40	0.381226092533246\\
41	0.380013951273471\\
42	0.37481490529554\\
43	0.378217963930757\\
44	0.374302124008181\\
45	0.368039475748562\\
46	0.364982082501503\\
47	0.363138684696915\\
48	0.364712345518167\\
49	0.364705577682637\\
50	0.362041769586605\\
51	0.362633088362811\\
};
\addlegendentry{Benchmark 1}
\addplot [color=black,line width=0.7pt]
  table[row sep=crcr]{%
1	0.344331978726371\\
2	0.333362980917311\\
3	0.345510523562089\\
4	0.33183558380407\\
5	0.410369416078303\\
6	0.361229428981142\\
7	0.482861694786393\\
8	0.374988108529712\\
9	0.476389428449323\\
10	0.33456115645152\\
11	0.485542921980078\\
12	0.411943140078269\\
13	0.568909358902835\\
14	0.203390843900455\\
15	0.431863700167206\\
16	0.319777502820628\\
17	0.683113306250815\\
18	0.569315022574662\\
19	0.653858370903611\\
20	0.50304081480298\\
21	0.725848597000986\\
22	0.450688210351872\\
23	0.755012020950157\\
24	0.40684057409937\\
25	0.686432764773835\\
26	0.29562806962129\\
27	0.779219364666216\\
28	0.26242336615861\\
29	0.827136116454939\\
30	0.27474573924351\\
31	0.966284927444233\\
32	0.321033476803637\\
33	0.944855691739333\\
34	0.340338684069967\\
35	1.00025410659248\\
36	0.246595890210914\\
37	0.389200570559332\\
38	0.224614062658963\\
39	0.381059295990961\\
40	0.258750038992228\\
41	0.376228937013893\\
42	0.295128809341324\\
43	0.37552140068457\\
44	0.26488224468313\\
45	0.366589120808727\\
46	0.301214857119466\\
47	0.364887785825838\\
48	0.358696213292598\\
49	0.359962722154455\\
50	0.340609141928655\\
51	0.360307437555218\\
};
\addlegendentry{Benchmark 2}

\addplot [color=purple, dashed, line width=1.0pt]
  table[row sep=crcr]{%
1	0.2\\
2	0.2\\
3	0.2\\
4	0.2\\
5	0.2\\
6	0.2\\
7	0.2\\
8	0.2\\
9	0.2\\
10	0.2\\
11	0.2\\
12	0.2\\
13	0.2\\
14	0.2\\
15	0.2\\
16	0.2\\
17	0.2\\
18	0.2\\
19	0.2\\
20	0.2\\
21	0.2\\
22	0.2\\
23	0.2\\
24	0.2\\
25	0.2\\
26	0.2\\
27	0.2\\
28	0.2\\
29	0.2\\
30	0.2\\
31	0.2\\
32	0.2\\
33	0.2\\
34	0.2\\
35	0.2\\
36	0.2\\
37	0.2\\
38	0.2\\
39	0.2\\
40	0.2\\
41	0.2\\
42	0.2\\
43	0.2\\
44	0.2\\
45	0.2\\
46	0.2\\
47	0.2\\
48	0.2\\
49	0.2\\
50	0.2\\
51	0.2\\
};
\addlegendentry{Threshold}
\end{axis}
\end{tikzpicture}%
}
	
		\caption{Achieved rate per grid}
		\label{BM}

\end{figure}

In this part we report the performance evaluation of our method in compared to benchmarks for the first scenario ($N_{\text{BS}}=64$, $N_{\text{UE}}=16$).
As the first benchmark, we consider the case where the paths skeleton will be updated at each grid. As it is shown in Fig. \ref{BM}, this benchmark has the highest performance with the cost of the higher query cost. 
As the second benchmark, we simulate the approach in \cite{8057188} which is based on updating the beamforming and combining vectors based on a fixed Euclidean distance. 
We choose the distance between two consecutive updates equal to $7$ m in order to keep the same (almost) total number of updates as our approach ($U_\text{Max}=20$). Fig. \ref{BM} shows the archived rate for the maximum value of the $r$ as the solution of \eqref{e}.
 The maximum value of the $r$ in this benchmark is $5$ m while our method with the same number of path skeleton updates can tolerate up to $r=10$ m. Moreover, high rate fluctuation in this benchmark can prohibit the support of a reliable connection to the user. The reason for those rate fluctuations is that the channel correlation is weak within the $7$ m distance and we need to update the beamforming and combining vectors meanwhile.

\section{Conclusions} \label{conclusions}

Localization information plays an important role in reducing the beamforming overhead in mmWave communications. In this work, we introduced a beamforming algorithm that can operate under imperfect location information and still provide a reliable connection. Our algorithm is based on tracking the spatial correlation of the path skeletons, i.e. available strongest path between the transmitter and the receiver. Our numerical results have shown different trade-offs between location information uncertainty and the achieved rate performance and the overhead. We also showed that wider beams can tolerate more location error with the cost of lower beamforming gain. 

\bibliographystyle{IEEEtran}
\bibliography{IEEEabrv,ref_bib}

\end{document}